\documentclass[runningheads]{llncs}

\usepackage{colortbl}
\usepackage{multirow}
\usepackage{subcaption}
\usepackage{tikz}
\usepackage{hyperref}

\usepackage{listings}
\usepackage[frozencache=true,cachedir=minted-cache]{minted}
\AtBeginEnvironment{minted}{%
  }

\newcommand{\circled}[2][]{
  \tikz[baseline=(char.base)]{%
    \node[anchor=text, shape=circle,draw, inner sep=0pt, minimum size=0.5em] (char){#1\strut};
    \node at (char.center) {\makebox[0pt][c]{#2}};}}
\robustify{\circled}
\usepackage[ruled, lined, linesnumbered, noend]{algorithm2e}
\DontPrintSemicolon
\SetAlFnt{\small}

\usepackage{etoolbox}
\makeatletter

\patchcmd{\@algocf@start}
  {-1.5em}
  {0pt}
  {}{}
\makeatother

\newcommand{\splits}[0]{{\sc SplITS}}
\newcommand{\fuzzware}[0]{{\sc Fuzzware}}
\newcommand{\icicle}[0]{{\sc Icicle}}
\newcommand{\ptwoim}[0]{{\sc P$^{2}$IM}}
\newcommand{\ember}[0]{{\sc Ember-IO}}

\begin{document}
\title{\splits{}: Split Input-to-State Mapping for Effective Firmware Fuzzing}

\author{Guy Farrelly\inst{1} \and
Paul Quirk\inst{2} \and
Salil S. Kanhere\inst{3} \and
Seyit Camtepe\inst{4} \and
Damith C. Ranasinghe\inst{1}
}
\authorrunning{G. Farrelly et al}
%

\institute{University of Adelaide, Australia
\email{\{guy.farrelly, damith.ranasighe\}@adelaide.edu.au} \and
Defence Science and Technology Group, Australia \email{paul.quirk@defence.gov.au} \and
University of New South Wales, Australia \email{salil.kanhere@unsw.edu.au} \and
CSIRO Data61, Australia
\email{seyit.camtepe@data61.csiro.au}
}

\maketitle              

\begin{abstract}
Ability to test firmware on embedded devices is critical to discovering vulnerabilities prior to their adversarial exploitation. State-of-the-art automated testing methods rehost firmware in emulators and attempt to facilitate inputs from a diversity of methods (interrupt driven, status polling) and a plethora of devices (such as modems and GPS units). Despite recent progress to tackle peripheral input generation challenges in rehosting, a firmware's expectation of multi-byte magic values supplied from peripheral inputs for string operations still pose a significant roadblock. 
We solve the impediment posed by multi-byte magic strings in monolithic firmware. 
We propose \textit{feedback mechanisms} for input-to-state mapping and retaining seeds for targeted replacement mutations with an efficient method to solve multi-byte comparisons. The feedback allows an \textit{efficient} search over a \textit{combinatorial solution-space}. 
We evaluate our prototype implementation, \splits{}, with a diverse set of 21 real-world monolithic firmware binaries used in prior works, and \textit{\textbf{3 new binaries}} from popular open source projects. \splits{} automatically solves 497\% more multi-byte magic strings guarding further execution to uncover new code and bugs compared to state-of-the-art. In 11 of the 12 real-world firmware binaries with string comparisons, including those \textit{extensively} analyzed by prior works, \splits{} outperformed, statistically significantly. We observed up to 161\% increase in blocks covered and discovered \textit{\textbf{6 new bugs}} that remained \textit{guarded} by string comparisons. Significantly, deep and difficult to reproduce bugs guarded by comparisons, identified in prior work, were found consistently. To facilitate future research in the field, we  release \splits{},
the new firmware data sets, and bug analysis at \textcolor{blue}{\url{https://github.com/SplITS-Fuzzer}}.

\end{abstract}
\section{Introduction}

Embedded device proliferation is creating new targets and opportunities for adversaries. 
Microcontrollers running firmware are becoming integral components of safety and security critical systems. In general, embedded devices take input from a diverse set of inputs and provide output in a unique manner, far different from those typically found on desktop computers. Integrated peripheral devices, such as timers or serial ports, manage communications with the user, often without any supervisory control from an operating system---\textit{importantly}, the lack of supervisory control reduces the ability to detect faults and abort~\cite{WYCINWYC}. But, security is a crucial enabler for connected devices, making scalable and automated methods to identify software bugs and vulnerabilities prior to public release a research and societal imperative. 

\begin{figure}[t!]
    \centering
    \includegraphics[width=0.75\linewidth]{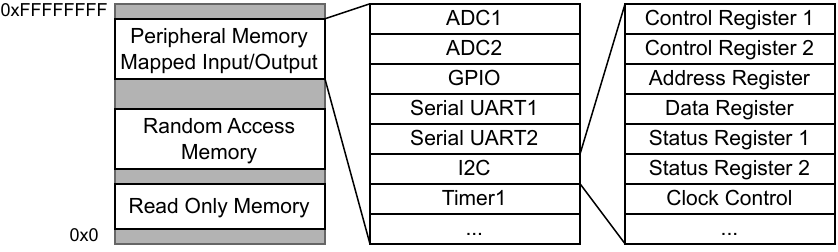}
    \caption{Example of peripherals in a microcontroller's memory address space. Each peripheral contains multiple memory mapped registers within the peripheral's Memory Mapped Input/Output (MMIO) region a firmware can interact with.}
    \label{fig:peripherals}
    \vspace{-2mm}
\end{figure}

Fuzz testing, or fuzzing, is a de-facto industry standard for software testing and can play a crucial role in developing secure connected devices through scalable and automated testing of firmware. In fuzzing, inputs are automatically generated and run to uncover unusual program behavior~\cite{aflplusplus}. But, the unique characteristics of embedded devices and their firmware present challenges for adopting fuzzing tools~\cite{WYCINWYC}. Fuzzing firmware based on their execution on physical devices~\cite{conware,avatar,avatar2,inception,pretender} is hampered by dependence on execution on low performance embedded processors limiting fuzzing performance. 
To improve fuzzing throughput, the firmware can be rehosted~\cite{sok_embedded} within an emulated environment on a high performance processor. While progress is made towards re-hosting and fuzzing Unix-based firmware~\cite{firmadyne,firmafl,firmae,firmfuzz,equafl}, fuzzing monolithic embedded firmware presents a unique set of challenges~\cite{WYCINWYC,p2im,ember,uEmu,fuzzware,icicle,HALucinator,Li_2021}. A problem for rehosting for fuzzing arises from developing methods to automatically provide peripheral inputs to the multitude of memory mapped interfaces a firmware may access on these devices as illustrated in Fig.~\ref{fig:peripherals}. In recent works~\cite{p2im,uEmu,fuzzware,ember,semu}, fuzzer generated test cases are fed to the framework rehosting the firmware, which uses this input stream to provide values to a set of peripheral registers, representing an interface used by a peripheral device (such as a modem or sensor). These approaches focus on ensuring data from peripherals can be successfully generated for the target.

\vspace{2mm}
\noindent\textbf{A Firmware Fuzzing Roadblock.~}Existing methods to automatically manage peripheral inputs do not consider the problem posed by multi-byte magic values expected in peripheral inputs by firmware binaries. We observe software used to communicate with external devices often depends on magic bytes. For example, external devices such as modems or GPS modules follow text based communications, and communicate these messages to the firmware using interfaces managed by on-chip peripherals. Reaching key code sections accessing these external peripherals, as illustrated in Listing~\ref{lst:stringcmp}, requires the delivery of magic bytes expected by the binary, via peripheral accesses.

\begin{listing}[b!]
    \centering
    \includegraphics[width=0.9\linewidth]{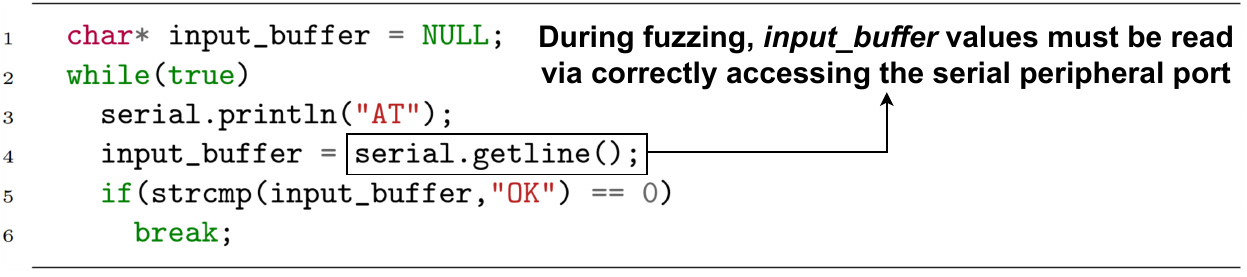}
    \caption{Simplified code snippet illustrating a wait for a peripheral device to respond \texttt{OK} to an AT command (extracted from \textbf{LiteOS IoT} binary). Failing to solve the string prevents execution of any deeper code, as the loop never exits.}
    \label{lst:stringcmp}
    \vspace{-2mm}
\end{listing}

The problems posed by magic-values are recognized as a difficult hurdle for fuzzers to overcome in general. For non-firmware binaries, techniques in ~\cite{redqueen,greyone,eclipser,laf_intel} are designed specifically to overcome these hurdles. The state-of-the-art technique for the problem in non-firmware binaries relies on \textit{input-to-state mapping}, where a sequence of bytes from the fuzzer generated input is mapped to a program state variable such as the \texttt{input\_buffer} in our example and are subsequently subjected to a surgical replacement by the magic values. Unfortunately, it cannot be simply employed for firmware.

Solving strings with input-to-state mapping techniques poses a unique challenge in firmware fuzzing due to the \textit{unusual} and \textit{diverse} input methods. While desktop applications read strings from a source, such as a file, as a contiguous block, embedded firmware processes \textbf{\textit{each}} input byte from a peripheral, one at a time, each requiring multiple registers within the peripheral to be read. Additionally, \textit{many peripherals are active simultaneously}, each with multiple memory mapped registers that influence firmware execution, as shown in Figure~\ref{fig:peripherals}. Between each byte of data read by a given peripheral, numerous other peripheral registers are often accessed, causing the data bytes within the string sourced from the fuzzer generated input to be spread across a region. Consequently, the fuzz inputs mapping to the expected magic bytes are \textit{unpredictably interspersed} throughout a large fuzz input. This prevents existing input-to-state mapping methods from identifying and solving string comparisons.

To address this gap, we consider a solution to the problem to advance recent progress made in monolithic firmware fuzzing to discover new software bugs previously guarded by magic-value roadblocks. 

\vspace{2mm}
\noindent\textbf{Our Approach.~}We propose new instrumentation and feedback to provide an effective method for input-to-state mapping and an optimized input replacement strategy to solve multi-byte string comparisons for fuzzing monolithic firmware. 

The feedback derived allows an \textit{efficient} search over the resulting \textit{combinatorial solution-space} by sequentially identifying the bytes in fuzz inputs corresponding to those used in comparisons. Subsequently, we attempt to simultaneously replace all bytes to ensure the multi-byte comparison is solved. Further, we exploit the nature of string representations to craft new instrumentation providing further feedback to the fuzzer to retain seeds for targeted replacement mutations. 
Together, these techniques create a robust method to overcome multi-byte magic strings by providing a faster exploration of the combinatorial search space of potential solutions for feedback driven fuzzing frameworks that \textit{simultaneously} provide input to many memory mapped peripherals.

\vspace{2mm}
\noindent\textbf{Our Contributions.~}We make the following contributions through this study:

\begin{itemize}
\setlength\itemsep{0.5em}
\item We propose an effective method to overcome hurdles posed by multi-byte string comparisons common in monolithic firmware.
\item
     To demonstrate the effectiveness of our techniques, we describe and build a prototype implementation, \splits{},  
     an effective method to consistently solve hard fuzzing problems dealing with multi-byte string comparisons without dependence on source code or debugging symbols for its implementation. 
\item
    In a series of extensive experiments with 24 real-world binaries, we demonstrate the method we propose is effective in significantly extending reachable code (up to 161\% increase in blocks covered with \splits{} automatically solving 497\% more multi-byte magic values) and discovering 6 previously unknown bugs (1 with a recent CVE assignment) guarded by string comparisons, compared to prior work; especially on binaries \textit{extensively analyzed by prior work}. We responsibly disclose vulnerabilities to the affected groups.

\end{itemize}

\urlstyle{tt}
To facilitate future research in the field, we will release \splits{},
the firmware data sets, and bug analysis at \textcolor{blue}{\url{https://github.com/SplITS-Fuzzer}}. We begin with a technical primer on monolithic firmware to help understand the unique aspects that present a challenge for fuzz testing monolithic firmware.

\section{Technical Primer}
\vspace{-1mm}
\noindent\textbf{Monolithic Embedded Systems.~}Unlike Unix-based firmware, a single monolithic program controls all aspects of a device's operation. The application code, system libraries and hardware abstractions are combined into a single binary sharing the same memory space. 
The firmware manages all functions and interacts with peripherals such as buttons or GPS units through Memory Mapped Input/Output (MMIO), hardware interrupts and Direct Memory Access (DMA).

\vspace{2mm}
\noindent\textbf{MMIO.~}
MMIO is a predefined segment of memory reserved for communicating directly with peripherals through reads and writes to the peripheral's registers. As shown in Figure~\ref{fig:peripherals}, each peripheral has a set of data, control and status registers. \textbf{Data registers} contain the data to be read or written. \textbf{Control registers} store the parameters for the peripheral's operation, such as speed or operating mode. \textbf{Status registers} describe the current state of the peripheral, indicating the presence of new data, or any errors.
For \textbf{Status} and \textbf{Control} registers, it is common for many individual bits within a register to have distinct purposes and many components of the peripheral's state are condensed to a few registers.
 
While peripherals such as timers and Analog to Digital Converters (ADCs) are embedded on-chip, not all peripherals are integrated directly into a microcontroller. Instead the microcontroller can be connected to other hardware components such as wireless modems. These external devices communicate with the microcontroller to provide data not observable with only the built in on-chip peripherals. To facilitate this communication, integrated peripherals are implemented to control data buses. Some examples are the Serial Peripheral Interface (SPI), Inter-Integrated Circuit (I2C) and Universal Asynchronous Receiver Transmitter (UART). In each case, the firmware accesses external peripherals through an MMIO interface exposed by these integrated peripherals, to read and write data to the bus.

\vspace{1mm}
\noindent\textbf{Hardware Interrupts.~}Peripherals also interact with firmware by generating interrupt signals, indicating an event has occurred. The CPU immediately jumps to an interrupt handler associated with the given signal, reads the peripheral status, and performs any necessary operations to handle this signal. Depending on the current status, handling the signal may involve reading new data, writing data to the peripheral, or discarding data containing errors. Afterward, the CPU returns to its prior location to continue its previous task.

\section{The Input-To-Sate Mapping Problem}
\vspace{-2mm}

\begin{figure}[t]
    \centering
    \includegraphics[width=\linewidth]{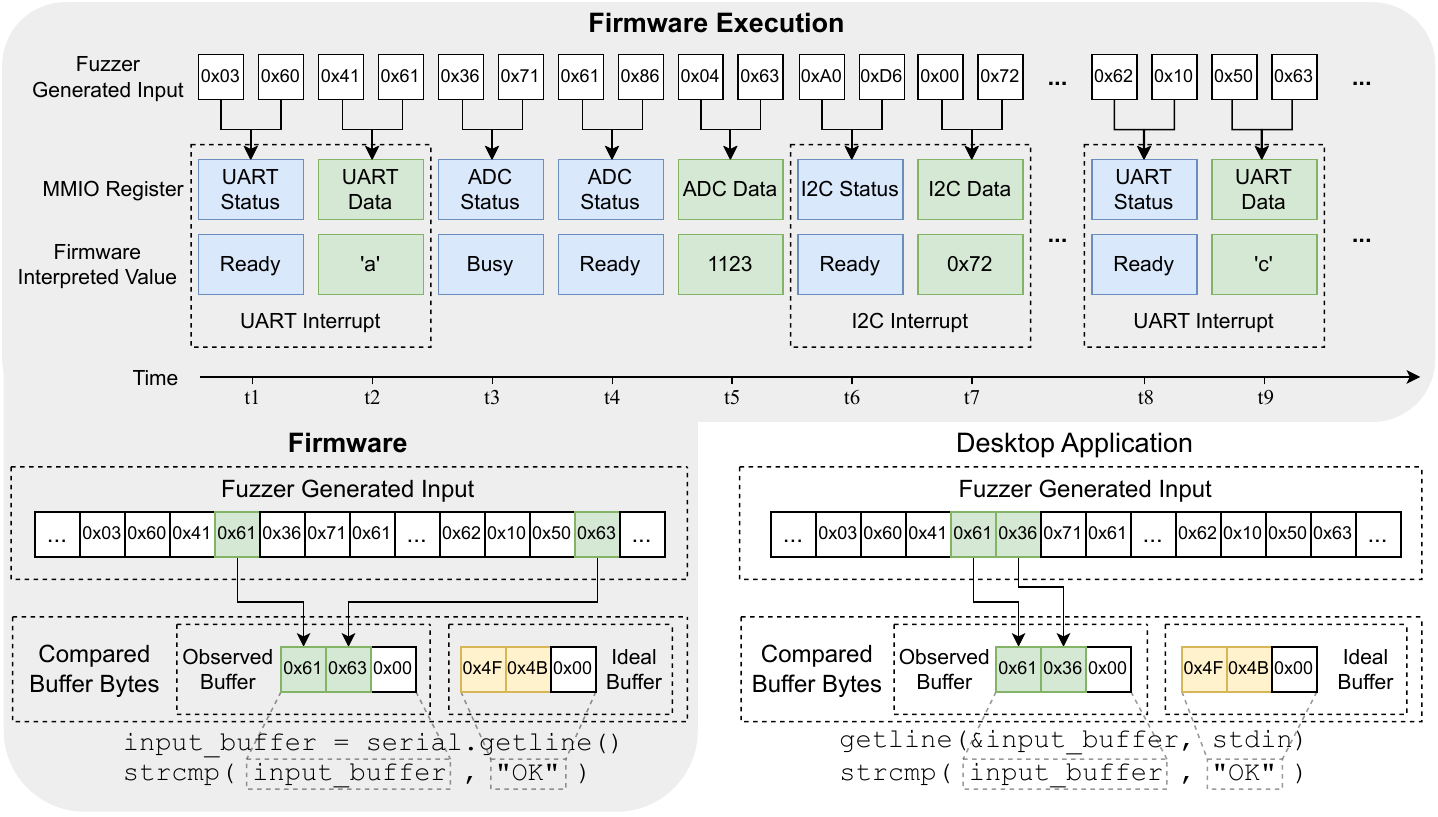}
    \caption{Excerpt of a fuzzer generated input consumed for a sequence of register reads for multiple active peripherals. The UART serial port loads the bytes \texttt{a} (\texttt{0x61}) at time \textit{t2} and \texttt{c} (\texttt{0x63}) at time \textit{t9}. The values read from the serial port are used to form the string \texttt{ac}. Although, adjacent in \textit{input\_buffer}, \texttt{a} and \texttt{c} are separated in the fuzzer generated input due to the unpredictable number of other peripheral register reads from \textit{t3} to \textit{t8}. In contrast, a desktop application loads the data in a block with the bytes adjacent in the input and \textit{input\_buffer}.}
    \label{fig:combinedaccess}
    \vspace{-3mm}
\end{figure}

The use of multi-byte magic values in the form of strings occurs regularly on embedded devices. These devices can use strings to take instructions on how they should operate, such as the use of GCode commands on a 3D Printer. Alternatively, firmware can communicate with external peripherals using strings. We observe this to be the case for AT commands used to communicate with modems, similar to the initialization example in Listing~\ref{lst:stringcmp}. Here, if the device does not respond with the string \texttt{OK}, the while loop will never exit, and the firmware will not proceed further. State-of-the-art methods use Input-To-State (ITS) mapping to overcome such hurdles. The goal is to solve this comparison by observing the value in \textit{input\_buffer} and locating it within the fuzzer generated test case, and replacing it with the ideal value \texttt{OK}. 

Existing input-to-state mapping techniques are effective for binaries where fuzzer generated data is loaded as a contiguous block for processing. But,  \textit{firmware typically loads fuzzer generated input data over peripherals over a period of time}, byte by byte with frequent switching between tasks for different peripherals, and uses multiple register accesses for each byte of input. This behavior \textbf{prevents the \textit{observed} data from being contiguous in the fuzz input}. Figure~\ref{fig:combinedaccess} shows an example of firmware using fuzzer generated input, prior to reaching the string comparison described in Listing~\ref{lst:stringcmp}, and a similar example for a desktop application. While the desktop application code loads the bytes 0x61, 0x36, corresponding to the ASCII string \texttt{a6}, the firmware example loads \texttt{ac} from a UART serial port through multiple MMIO register reads, intermitted by other peripheral accesses. While the string \texttt{a6} does appear in the fuzzer generated input, the string \texttt{ac} does not. Consequently, mapping the \textit{observed} value back to the fuzzer generated input is not trivial. This prevents existing
input-to-state mapping methods from identifying and solving string comparisons.

To address this gap, we develop an input-to-state mapping technique for monolithic firmware. 
Through our efforts, we aim to advance recent progress made in monolithic firmware fuzzing to discover new software bugs
previously guarded by magic-value roadblocks.

\section{Split Input-to-State Mapping}
\vspace{-2mm}
Consider Fig.~\ref{fig:combinedaccess}. A naive approach to solve the non-contiguous string is to \textit{attempt all possible combinations} of replacements for each character in the \textit{observed string} with the corresponding character in the \textit{ideal string}. While the approach may be possible for the example in Listing~\ref{lst:stringcmp}, where the string comparison only depends on solving two characters, 
this solution does not scale, instead creating a combinatorial problem. The number of executions required to test all possible combinations depends on two factors: \textit{the length of the string} and \textit{the number of candidate bytes for replacement} in the fuzzer generated input.  
Notably, we observe strings up to 18 characters in length, and fuzzer generated inputs with thousands of bytes. These large inputs and strings lead to creating a large search space infeasible to test effectively, as shown in the following examples.

\vspace{1mm}
\noindent\textbf{\textit{Example~1:} \textit{Long Strings.~}} \textit{Consider the string,} \texttt{rpl-refresh-routes} \textit{observed in the} \textbf{Contiki-NG} \textit{binary. When three candidates for each of the 18 characters are considered, the number of possible combinations rapidly explodes. This string would require more than three hundred million combinations be tested.}

\vspace{1mm}
\noindent\textbf{\textit{Example~2:} \textit{Many Candidate Bytes.~}} \textit{Consider the string} \texttt{poweron} \textit{observed in the} \textbf{Console} \textit{binary. When 16 candidates for each of the seven characters are considered, more than two hundred million combinations would be tested.}

To illustrate how the naive approach is influenced by the inclusion of longer strings and more candidates for each character, we refer to the naive search \circled{1} in Figure~\ref{fig:replacements}. This example uses a four character string, each with four occurrences in the fuzzer generated input that are candidates for replacement. This results in 256 possible combinations. Each additional character has a multiplicative impact on the number of combinations to test. The number of occurrences of this additional character in the fuzz input defines the scale of this multiplicative impact. While certain constraints can be implied, such as those discussed in Section~\ref{sec:opt}, to remove some combinations, a naive approach is impractical and highly inefficient, as it does not prevent the extreme growth in search space resulting from this method.

\subsection{Feedback Guided Search}\label{sec:identify_bytes}
Rather than considering the possible combinations of replacements, we consider a more scalable approach guided by  \textit{whether a given byte influences the string comparison}.
\begin{figure}[t!]
    \centering
    \includegraphics[width=0.94\linewidth]{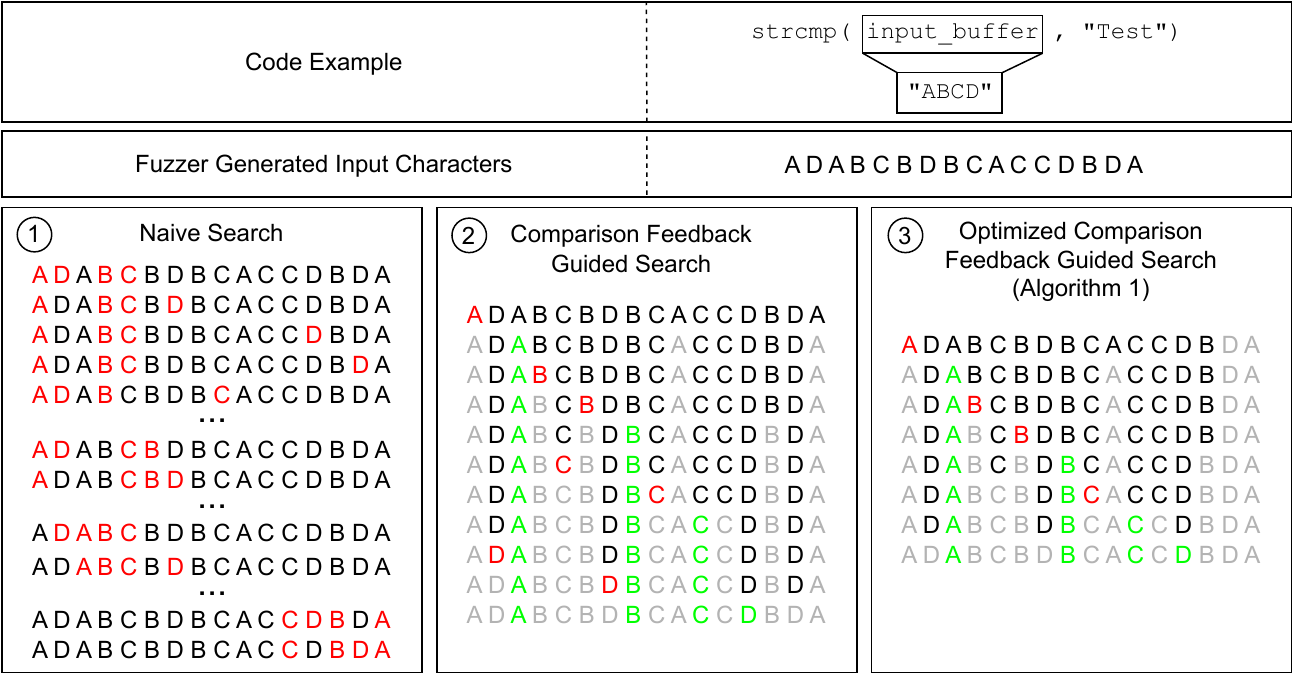}
    \caption{Example of naive vs. feedback guided search of the string ``ABCD" within a fuzzer generated input consisting of four As, Bs, Cs and Ds. In \circled{1}, the four occurrences of the character `D', creates four additional cases to test \textbf{for each combination} of the string ``ABC", causing the search space to be four times larger than searching for the string ``ABC" within the input. The same addition for the feedback guided searches \circled{2} \& \circled{3} only adds, at most, four more tests.}
    \label{fig:replacements}
\end{figure}

\vspace{2mm}
\noindent\textbf{Comparison Feedback.~}To successfully replace memory comparisons based on non-contiguous accesses, without introducing an explosion in candidate replacements, we attempt to solve each byte individually. We propose monitoring the data present in the \textit{observed} and \textit{ideal} value buffers, and \textbf{\textit{use the content of the buffers as feedback mechanism}}. If the current byte has successfully been updated within the \textit{observed buffer}, we 
only consider combinations that include this byte. If this value remains unchanged, or the comparison is not reached, we proceed to the next candidate location within the input and no longer consider any replacements that include this byte. This is done until all bytes in the \textit{observed buffer} match the byte at the corresponding index in the \textit{ideal buffer}, or an \textit{observed buffer} byte could not be located in the input.

Figure~\ref{fig:replacements} \circled{2} shows the feedback guided search in an example, progressively locating the 'A', 'B', 'C' and 'D' characters within an input that loads \texttt{ABCD} into a buffer. Bytes that have been confirmed to be a part of the observed buffer are highlighted in green. Tested bytes that do \textbf{not} correspond to a byte in the \textit{observed buffer} are shown in red; such as the first  'A' tested once before exclusion. Additionally, the fourth 'A' character is quickly discarded from the search space (shown in gray), as a valid replacement for this character in the \textit{observed buffer} has already been identified. When compared to the naive search, the number of firmware executions required to solve the string comparison is greatly reduced. As each candidate byte in the fuzzer generated input can only be considered for replacement once, \textit{this method only has linear complexity with regard to the string length, and number of candidates for each character}.

\vspace{2mm}
\noindent\textbf{Length Feedback.~}\label{length}
Comparison feedback alone is not adequate for an effective search and replacement strategy. \textit{Strings must be the same length to be equal}. In cases where the \textit{observed string}---values generated by the fuzzer for a peripheral---is shorter than the \textit{ideal string}, we do not have any knowledge of which bytes need to be mutated to extend the \textit{observed string} to the required length. But, during fuzzing, we can expect many executions to contain different strings of varying lengths. However, due to the coverage metrics used by greybox fuzzers, if the inputs only reach the same code, they are rarely saved for further processing. 
Thus, we rarely recall seeds with long enough strings to attempt replacement of the \textit{observed string}. To address this issue, we include additional coverage instrumentation. 
We exploit the null byte in string formats to determine their length. By searching for null terminators in the \textit{observed} and \textit{ideal} buffers, we can determine the length of the \textit{observed string} and \textit{ideal string}.
When a string comparison is identified, we ascertain the string lengths, and count previously unseen \textit{observed string} lengths similar to the length of the \textit{ideal string} as interesting.  
This ensures we save inputs that load sufficient data into the \textit{observed string buffer} for successful replacement with the \textit{ideal string}.

\begin{algorithm}[b!]
\caption{Feedback Guided Search \& Replacement Algorithm for solving a string comparison. A comparison identifiable by \textit{cmp\_id}, reached when executing the firmware with \textit{input}, requires the \textit{observed\_string} to match the \textit{ideal\_string}. \textit{GetReadBytes} returns the number of input bytes read at the time of comparison, and \textit{InsertDelimiter} replaces a byte in a string with common tokenization delimiters.}\label{fig:splits_alg}
\DontPrintSemicolon
\SetKwInOut{Input}{input}
\SetKwFunction{execute}{Execute}
\SetKwFunction{getobserved}{GetObserved}
\SetKwFunction{getreadbytes}{GetReadBytes}
\SetKwFunction{insertdelimiter}{InsertDelimiter}
\Input{$input$, $ideal\_string$, $observed\_string$, $cmp\_id$}
$original\_input \gets input$\;
$last\_index \gets 0$\;
\tcc{For each byte value to be replaced}
\For{$i = 0~\KwTo~observed\_string.length$}{
    \For{$j~=~last\_index~\KwTo~\getreadbytes(cmd\_id)$}{
        \tcc{If this input byte is a candidate for replacement}
        \If{$observed\_string_i = input_j$}{
            $input_j \gets ideal\_string_i$\;
            $\execute{input}$\;
            $new\_observed \gets \getobserved{cmp\_id}$\;
            \eIf{$new\_observed.exists$~{\rm {\bf and}}$~new\_observed_i = ideal\_string_i$}{
                $last\_index \gets j + 1$\;
                \textbf{break }\tcc{Solved. Move to next byte.}
            }{
                $input_j \gets original\_input_j$ \tcc{Not solved. Restore byte}
            }\If{$j = \getreadbytes{cmp\_id}$}{
                \KwRet \tcc{Unable to map byte}
            }
        }
    }
    \If{$i >= ideal\_string.length$}{
        $\insertdelimiter{observed\_string, i}$ \tcc{Shorten observed string}
        \KwRet
    }
}
\end{algorithm}
\subsection{Search Optimizations}\label{sec:opt}
To further minimize the effort required to discover useful inputs, we consider several \textit{intuitive} but \textit{effective} optimizations to reduce the number of combinations considered and improve applicability. These optimizations are combined with the comparison feedback described in Section~\ref{sec:identify_bytes} to develop Algorithm~\ref{fig:splits_alg}.

\vspace{1mm}
\noindent\textbf{Constrained Mutable Regions.~}We constrain the sections of the input considered valid for mutation. We expect  any data that has not been read prior to the comparison, cannot influence the result of the comparison. Consequently,  we do not attempt to mutate these bytes. This reduces the maximum number of iterations of the inner loop in Algorithm~\ref{fig:splits_alg}.

\vspace{1mm}
\noindent\textbf{Implied Sequential Access.~}In the context of firmware, we can assume that data, whilst not adjacent, is read in sequence (i.e. read of the second byte in the string occurs a period after the first) due to the stream-like nature of the input. Consequently, for all \textit{observed buffer} bytes after the first, we exclude any input bytes that have not been read since the previously solved byte in the \textit{observed buffer} was accessed. This reduces the range of \textit{j} in the inner loop of Algorithm~\ref{fig:splits_alg}.

\vspace{1mm}
\noindent\textbf{String Contractions.~}~For strings to be equivalent, both strings must be the same length. While we can not extend the \textit{observed string}, we can attempt to shorten it. When the \textit{observed string} is longer than the ideal string, we identify the byte corresponding to the next character in the \textit{observed string} and attempt replacement of this byte with a set of common token delimiters such as space and newline, contracting the string to the correct size.

In Figure~\ref{fig:replacements}, we illustrate the impact of using the optimizations in block \circled{3}. The optimizations reduce the number of executions required to identify the bytes in the comparison compared to the search in \circled{2} guided solely by feedback.

\section{Implementation}

\begin{figure}[t!]
    \centering
    \includegraphics[width=0.9\linewidth]{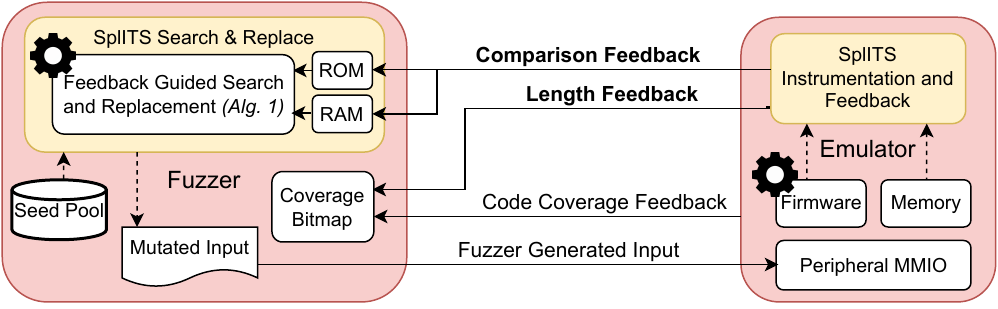}
    \caption{Overview of \splits{} design. The Instrumentation and Feedback component provides string comparison and length feedback. Length feedback is implemented by setting bits within the existing fuzzer coverage bitmap (Section~\ref{length}). The compared strings and information on which bytes of the mutated input have been read are provided to the \splits{} Search \& Replace module, where the comparison is solved, guided by incremental feedback  (Algorithm~\ref{fig:splits_alg}).}
    \label{fig:systemoverview}
\end{figure}

We provide an overview of a realization of our techniques for a monolithic firmware fuzzer in Figure~\ref{fig:systemoverview}. To test the efficacy of our approach, we implemented our method with \fuzzware{}~\cite{fuzzware}. We also apply the FERMCov technique described in \textsc{Ember-IO}~\cite{ember} as it can increase the effectiveness of the framework. 
In particular, \fuzzware{'s} emulator was modified to detect potential string comparison functions, and report the location and buffer contents for each of these function calls to the fuzzer, AFL++~\cite{aflplusplus}. We implement our string replacement on top of AFL++'s existing \textit{cmplog} interface.  
AFL++ was configured to use random bytes for colorization, and only consider string comparisons for input-to-state solving.

Given that embedded system fuzzing frameworks such as \fuzzware{} use the AFL++ generated input as a stream, where each byte is only read once and in order, we use the cursor location within the stream to determine which data has been read at a given time. All bytes at a lower index than the current cursor have been read, and equal or higher indexes are unread. We pass the cursor index at the time of comparison from the \splits{} feedback inside the emulator back to AFL++ as an extension to the \textit{cmplog} interface. This knowledge of unread bytes is used as described in Section~\ref{sec:opt}. The index of the previously solved byte is determined internal to AFL++ during \splits{'} replacements.

Similar to the original \textsc{RedQueen}~\cite{redqueen} implementation, we detect comparison functions by hooking instructions used to call functions, and analyzing the parameters. In the case of ARM, we hook the branch linked (BL) instruction, and read the R0 and R1 registers, corresponding to the first two function parameters. A function is considered a potential memory comparison function when the first two parameters are valid memory pointers, where one points to ROM (assumed to be the \textit{ideal string}), and the other to RAM (assumed to be the \textit{observed string}). Our implementation does not require modifications to the firmware's source code or binaries, nor does it depend on debug symbols being present.

To handle nested substring checks within the same string, \splits{} is configured to attempt replacement at both the start and end of a string buffer, provided the current \textit{observed string} is longer than the \textit{ideal string}.

\section{Evaluation}

To determine \splits{'} effectiveness at solving non-contiguous string comparisons, we evaluate it on a set of monolithic real-world firmware. We analyzed the binaries used in evaluations from existing firmware fuzzing frameworks \ptwoim{}~\cite{p2im}, \textmu{}Emu~ \cite{uEmu}, \fuzzware{}~\cite{fuzzware} and \textsc{Ember-IO}~\cite{ember} that emulate memory mapped IO. From the 21 real-world firmware used in the coverage comparisons, nine contain functionality guarded by string comparison functions such as \textit{strcmp}, \textit{strncmp} and \textit{strstr}.  
We additionally included three \textit{new} targets containing string comparisons. Each binary is fuzzed for 24 hours in five trials. To ease the analysis, we considered the binaries in the following groupings:

\begin{itemize}
\item \textit{Magic-String-Binaries.~}We examined a set of 24 real-world binaries and observed magic values used in string comparisons in 12 of the binaries. We employed this set for our extensive evaluations. 
\item \textit{Other-Binaries.~}The remaining 12 real-world binaries that did not make use of any string comparisons to guard code execution.
\end{itemize}

We conduct evaluations \textit{with} and \textit{without} \splits{}. Here, the fuzzer without our techniques is the state-of-the-art \fuzzware{} improved with FERMCov (denoted simply as \fuzzware{}) while the \splits{} denoted implementation is \fuzzware{} improved with FERMCov along with our proposed techniques. \fuzzware{}~\cite{fuzzware} has previously shown higher code coverage and bug finding performance when compared to prior works such as \ptwoim{}~\cite{p2im} and \textmu{Emu}~\cite{uEmu}. \icicle{} with CompCov is included for a point of comparison as CompCov instrumentation could solve some string comparisons. We also included the recent state-of-the art monolithic firmware fuzzing framework, \ember{}. Tests were performed using an AMD Threadripper 3990X. We used the gathered coverage information and crashing inputs to answer the following questions:

\begin{enumerate}
    \item How successful is \splits{} at solving string comparisons?
    (Section~\ref{sec:eval_solving})
    \item Does the solving of string comparisons provide a significant increase in reachable code?
    (Section~\ref{sec:codecov})
    \item Does \splits{} impact fuzzing firmware with no comparisons?
    (Section~\ref{sec:overhead})
    \item How does \splits{} impact the discovery of firmware bugs?
    (Section~\ref{sec:bugs})
\end{enumerate}

\subsection{Effectiveness at Solving Strings}\label{sec:eval_solving}

Table~\ref{fig:string_summary} shows the ability of \fuzzware{}, \splits{} and \icicle{} with CompCov to solve multi-byte string comparisons that guard further code exploration.  
The results show using \fuzzware{} alone struggled to solve the majority of the string comparisons. For those that were solved, the ability to solve a given string was rarely consistent across the five trials. \icicle{} with CompCov was able to solve some strings unreachable by \fuzzware{} in the 3D Printer, Console and Zephyr SocketCan binaries. 
In contrast, \splits{} was able to perform significantly better and more consistently; solving all of the same strings across all trials for eight of the twelve binaries tested (i.e. the \textit{Min}  and \textit{Total} results were equal). In the remaining cases, the majority of strings were consistently solved, but a handful were not consistent as additional constraints prevented the fuzzer from reaching the string comparison. These constraints included the number of parameters given after a command, or other aspects of the system state that must be correctly set in conjunction with the provided string.

\begin{table}[t]
\caption{The number of reached string comparisons guarding code solved across five 24 hour trials in the \textit{Magic-String-Binaries}. \textit{Total} represents the number of unique comparisons solved at least once across all of the trials. \splits{} outperforms state-of-the-art fuzzing techniques and is demonstrably highly effective and yield \textit{consistent} performance across repeated runs.}
\label{fig:string_summary}
\centering
\resizebox{0.95\linewidth}{!}{
\begin{tabular}{|l|l|l|l|l||l|l|l|l||l|l|l|l|}
\cline{2-13}
\multicolumn{1}{c|}{} & \multicolumn{4}{c||}{\textsc{Fuzzware}}  & \multicolumn{4}{c||}{\textsc{ICICLE} (CompCov)} & \multicolumn{4}{c|}{\textsc{\splits{}}}\\
\hline
\textbf{Firmware}          & \textbf{Min}      & \textbf{Med} & \textbf{Max} & \textbf{Total} & \textbf{Min}                  & \textbf{Med}      & \textbf{Max}         & \textbf{Total} & \textbf{Min}                  & \textbf{Med}      & \textbf{Max}         & \textbf{Total}\\
\hline
\hline
3D Printer                     & 0        & 0      & 0   & 0     & 0        & 0      & 1   & 1 & \textbf{4}    & \textbf{4} & \textbf{4} & \textbf{4}    \\
\hline
Console                          & 1        & 1      & 2   & 2  & 2        & 3      & 6   & 9   & \textbf{15}    & \textbf{15} & \textbf{15} & \textbf{15}    \\
\hline
GPS Tracker                       & 0        & 0      & 0   & 0   & 0        & 0      & 0   & 0  & \textbf{4}    & \textbf{4} & \textbf{4} & \textbf{4}    \\
\hline
LiteOS IoT                        & 0        & 0      & \textbf{1} & \textbf{1}        & 0      & 0   & \textbf{1}  & \textbf{1}     & \textbf{1}    & \textbf{1} & \textbf{1} & \textbf{1}    \\
\hline
RF Door Lock                      & 0        & 0      & \textbf{1}        & \textbf{1}   & 0   & \textbf{1}   & \textbf{1}  & \textbf{1}     & \textbf{1}    & \textbf{1} & \textbf{1} & \textbf{1}    \\
\hline
Steering Control                  & 0        & 0      & 0   & 0  & 0        & 0      & 0   & 0   & \textbf{2}    & \textbf{2} & \textbf{2} & \textbf{2}    \\
\hline
uTasker MODBUS                   & 0        & 0      & 0   & 0  & 0        & 0      & 0   & 0   & \textbf{44}    & \textbf{45} & \textbf{45} & \textbf{45}     \\
\hline
uTasker USB                    &    0      &    0  &  0  &   0  & 0        & 0      & 0   & 0 & \textbf{31}     & \textbf{33} & \textbf{34} & \textbf{34}    \\
\hline
Zephyr SocketCan                & 2        &    26  & 30  & 30  & 29        & 32      & 33   & 40  & \textbf{52}    & \textbf{55} & \textbf{60} & \textbf{65}    \\
\hline
ChibiOS RTC                 & 0        &    0  & 1  & 2  & 0        & 0      & 2   & 2  & \textbf{14}    & \textbf{14} & \textbf{14} & \textbf{14}    \\
\hline
Contiki-NG Shell                & 0        &    0  & 0  & 0  & 0        & 0      & 0   & 0  & \textbf{14}    & \textbf{14} & \textbf{14} & \textbf{14}    \\
\hline
RiotOS TWR                 & 0        &   0  & 0  & 0   & 0        & 0      & 0   & 0 & \textbf{13}    & \textbf{13} & \textbf{16} & \textbf{16}    \\
\hline
\end{tabular}
}
\end{table}

\vspace{2mm}
\noindent\textbf{Performance Analysis.~}For a deeper understanding of the effectiveness of our approach, we evaluated the efficiency with which comparisons are solved by \splits{} compared to the state-of-the-art in \textbf{Appendix}~\ref{apd:deep-dive}

\subsection{Code Coverage Analysis}\label{sec:codecov}

We analyze code coverage over the two categories of firmware we devised (\textit{Magic-String-Binaries} and \textit{Other-Binaries}) to understand the effectiveness of \splits{} and determine if the introduction of the techniques impacts fuzzing performance. 

\begin{figure}[h]
    \centering
    \includegraphics[width=\linewidth,trim={0cm 0.3cm 0cm 0.3cm},clip]{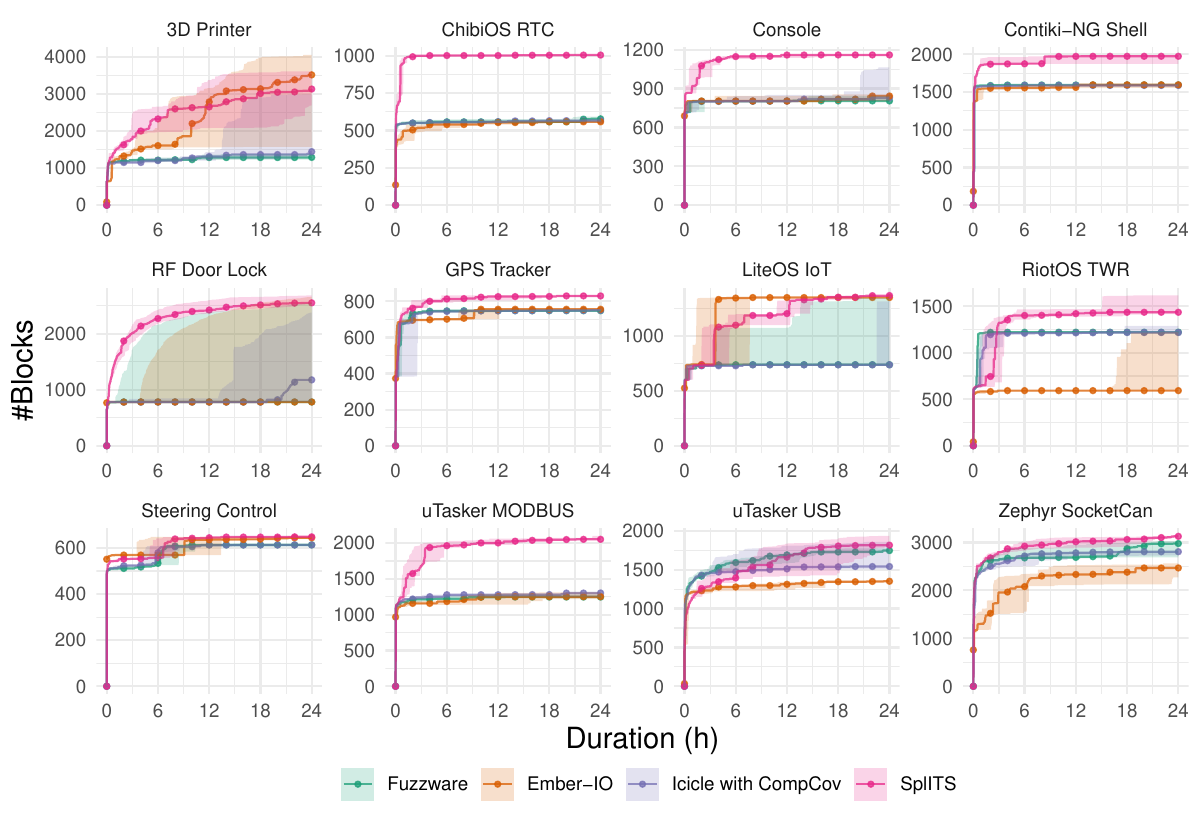}
    \caption{\textit{Magic-String-Binaries.} Comparison of coverage achieved by: {\sc Fuzzware}, \splits{}, \ember{} and \icicle{} with CompCov for \textit{Magic-String-Binaries}. Bands represent the range of coverage observed over five 24 hour trials.}
    \label{fig:cov_plot}
\end{figure}

The introduction of \splits{} has a large impact on code coverage for several of the firmware, shown in Figure~\ref{fig:cov_plot} and \textbf{Appendix}~\ref{apd:coverage}. Of the twelve Magic-String-Binaries, eleven had higher results deemed statistically significant when applying a Mann-Whitney U test at the 0.01 significance level. When compared to other frameworks, \splits{} saw an increase deemed statistically significant in 9 and 10 of the twelve binaries for \ember{} and \icicle{} respectively.

Compared to the other tested firmware fuzzing frameworks, \splits{} saw more than a 50\% increase in code coverage in both the uTasker MODBUS and the ChibiOS RTC binaries. We observed improvements upwards of 37\% for the Console binary, 23\% for the Contiki-NG Shell binary, and 17\% for the RiotOS TWR binary in median blocks covered when compared to the other frameworks. The GPS Tracker and Zephyr SocketCan binaries each saw an increase in median code coverage between 5 and 10\% compared to the next highest framework.

The string comparisons guarding initialization for the RF Door Lock and LiteOS IoT binaries are observable in the coverage results. 
\fuzzware{} could not consistently find a valid solution in 24 hours. This led to much higher minimum and median coverage results for \splits{}, an increase of more than 80\% in each of these cases. Due to \fuzzware{} and \ember{} both solving the string comparison in the RF Door Lock in a single trial, statistical significance could not be established. In contrast, \icicle{} with CompCov allowed solving the RF Door Lock binary's string comparison in the majority of tests. Interestingly, \ember{} was able to consistently find an input passing the two character string comparison found in the LiteOS IoT binary.

Block coverage in the 3D Printer increases more than 100\% when \splits{} is applied compared to \fuzzware{}. Most of the commands given to the 3D Printer firmware are parsed byte by byte, rather than as a string. The string comparisons come from a handful of exceptions, such as the emergency stop command \texttt{M112}. The number of code blocks reachable from these special cases is too small to account for the change increase in coverage from enabling \splits{}. Further investigation reveals that solving a single command can provide a mutable case with the correct format. After the \texttt{M112} emergency stop command was solved with \splits{}, mutation yielded discovery of the similar \texttt{M111} and \texttt{M113} commands. \ember{} was also able to uncover many of these commands.

Given the large difference in the number of solved strings between \fuzzware{} and \splits{} on the uTasker USB firmware, we consider the difference in the number of blocks reached comparatively small. Further investigation revealed that \splits{} and \fuzzware{} explored different parts of the firmware. Of the 2266 unique blocks executed across all runs, 226 were only reached using \fuzzware{}, while 437 blocks could only be reached when \splits{} was applied.

\vspace{1mm}
\noindent\textbf{Code Coverage of Other-Binaries.~}\label{sec:overhead}\splits{} is intended to solve strings, the attempts to solve strings require firmware executions to both search for, and attempt to solve potential string comparisons. To determine whether the additional processing impacts fuzzing of firmware with no string comparisons, we tested the \textit{Other-Binaries} and compare the coverage results to \fuzzware{} without \splits{} applied. We defer the results to \textbf{Appendix}~\ref{apd:coverage}. In summary, none of the tests revealed a statistically significant difference in coverage achieved; this suggests that fuzzing performance is not impacted by \splits{}.

\subsection{Bug Discovery}\label{sec:bugs}
While code coverage is a useful metric for comparing the performance of fuzzing frameworks, fuzzing aims to identify bugs. We deduplicated the reported crashes and investigated any crashes in these binaries identified with \splits{} that were not previously found and reported by \fuzzware{}.

\vspace{2mm}
\noindent\textbf{Analysis of Newly Discovered Bugs.~}\label{sec:new-bugs}After triaging the deduplicated crashes, we identified the following new firmware bugs, and verified they are not false positives. We responsibly disclosed any vulnerabilities to the appropriate vendor.

\vspace{1mm}
\noindent\textbf{(1)~Zephyr SocketCan.~}\textit{CVE-2023-0779.} Within Zephyr's network component, we identified a command used to provide information about internal packet buffers. The function takes a parameter in the form of an address pointer, directly from the user. This pointer is not validated beyond checking if it is null, allowing a user to point at any location in memory.

\vspace{1mm}
\noindent\textbf{(2)~Console.~}Within the Console binary, we identified an out-of-bounds read bug. Commands take input from the user regarding dates, without validation. Entering months greater than 12 causes data from outside an array to be read. We note this bug was concurrently discovered in \icicle{}~\cite{icicle}, when used with CompCov~\cite{aflplusplus}, which we replicated in one of our five trials.

\vspace{1mm}
\noindent\textbf{(3)~uTasker MODBUS \& uTasker USB.~}The uTasker binaries previously tested in \fuzzware{}~\cite{fuzzware} were compiled with the \textit{MEMORY\_DEBUGGER} enabled. This allows reading and writing arbitrary memory. While this is a valid method to crash the firmware, and ideally attempts to access unmapped memory would be restricted, we do not consider this a vulnerability, as these commands are expected to be disabled in release versions of firmware.

\vspace{1mm}
\noindent\textbf{(4)~Contiki-NG Shell.~}The Contiki-NG Shell binary does not always validate the number of parameters given to a shell command. Attempting to call a vulnerable command with fewer parameters causes the parameter pointer to be set to null. This null pointer is not checked prior to being dereferenced.

\vspace{1mm}
\noindent\textbf{(5-6)~GPS Tracker.~} We uncovered two new crashes, both null pointer dereferences caused by a lack of error checking. For the first bug, the code assumes a call to \textit{strtok} will always return at least one token, and fails to consider the case where the input is an empty string. 
The second bug fails to consider a received string may not contain an expected value. Searching for the expected substring results in a null pointer being returned when it is not found, then dereferenced. This bug is identical in nature to a known bug in this firmware, and appears to be caused by copy-pasting the flawed code to other sections of the firmware.

\vspace{2mm}
\noindent\textbf{Analysis of Existing Bugs Guarded by String Comparisons.}\label{sec:known-bugs}
Notably, the inclusion of \splits{} does not change the execution of a given fuzzer generated input. Other mutation stages such as \textit{havoc} and \textit{splice} employed by AFL++ can still produce inputs that trigger the crashes previously reported by \fuzzware{}. We observe two bugs previously reported by \fuzzware{} are guarded by string comparisons. The GPS Tracker contains a null pointer dereference when parsing AT commands. In our tests, none of the five \fuzzware{} trials triggered this bug. 
The RF Door Lock contains a bug in the main loop, requiring the string comparison in initialization to be solved. This deeper bug was triggered in one of our five \fuzzware{} trials, as the others were unable to proceed past initialization. \splits{}, was able to reproduce both bugs in \textbf{all} five trials.

\section{Discussion}\label{sec:discussion}
\noindent\textbf{Comparison Functions.~}Predominantly, we observed standard C functions for comparing strings. But, in the two uTasker binaries, a custom function performs string comparisons, \textit{fnCheckInput}. However, the custom function is detected as a string comparison function and appropriately solved. This suggests \splits{} can be applicable outside of the standard C functions. Notably, the use of functions to perform comparisons is extremely common but it could be performed inline, preventing the string comparison from being detected and solved using \splits{}. However, we have not observed this problem in our testing.

\vspace{1mm}
\noindent\textbf{Comparison Detections.~}We observed some cases of false positive detection of string comparison functions. One example is printing to a serial port. The static string to be printed is detected as an \textit{ideal string}, while the pointer to the serial device is detected as an \textit{observed buffer}. This has little significance. Because the only impact from these false positives is taking extra computation time but time loss is already minimized, as we do not expect to be able to map the input to many of the serial port's internal variables. If a byte cannot be mapped, further processing of this comparison is abandoned.

\vspace{1mm}
\noindent\textbf{Applicability to Other Frameworks.~}\label{other_frameworks}While we have implemented \splits{} on top of \fuzzware{}~\cite{fuzzware}, it is also applicable to other monolithic embedded firmware fuzzing frameworks such as \ptwoim{}~\cite{p2im}, \textmu{Emu}~\cite{uEmu} and Ember-IO~\cite{ember}. In the case of Ember-IO, there is an additional requirement that \textit{peripheral input playback} was not the source of any of the bytes in the \textit{observed value} buffer; as only fuzzer generated input bytes can be located using input-to-state mapping. We consider DMA out of the scope, and the application of \splits{} to works that handle DMA such as DICE~\cite{dice} and SEmu~\cite{semu} has not been considered.

\section{Related Work}
\vspace{-3mm}
\noindent\textbf{Embedded Firmware Fuzzing.~}In general, for Unix-based firmware, several approaches have been developed~\cite{firmadyne,firmafl,firmae,firmfuzz,equafl}. While acknowledging these efforts, in this study, we focus on fuzzing approaches for monolithic firmware.

\vspace{1mm}
\noindent\textbf{Monolithic Embedded Firmware Fuzzing.~}To function on monolithic firmware, some works replace hardware interactions at a Hardware Abstraction Layer~\cite{Li_2021,HALucinator}, assisted by manual effort. To automate this process, works such as \ptwoim{}~\cite{p2im} and \textsc{\textmu{Emu}}~\cite{uEmu} use heuristic based models to classify individual registers and infer the behavior of each classification. 
Several works have explored the use of symbolic execution as a tool to assist in firmware testing~\cite{uEmu,FIE,jetset,laelaps,mousse}. \textsc{SEmu}~\cite{semu} instead models the behavior of peripheral registers by processing the manufacturer's manuals. Approaches such as \textsc{Ember-IO}~\cite{ember} and \fuzzware{}~\cite{fuzzware} fuzz all registers, and simplify the process by reducing the amount of mutation required.

\vspace{1mm}
\noindent\textbf{Input-to-State Correspondence Methods.~}Early methods to assist multi-byte comparisons such as \textsc{laf-intel}~\cite{laf_intel} and CompCov~\cite{aflplusplus} replace these comparisons with a series of easier to solve single byte comparisons. Alternate approaches used symbolic execution~\cite{klee,driller} or taint tracking~\cite{vuzzer,angora,taintscope,greyone} as a means to uncover paths with difficult to solve constraints. Eclipser~\cite{eclipser}, \textsc{RedQueen}~\cite{redqueen} and \textsc{WEIZZ}~\cite{weizz} use input-to-state mapping for non-firmware binaries to inform replacement mutations. These desktop-focused approaches assume contiguity between program state and inputs when solving string comparisons.

\section{Conclusion}
\vspace{-3mm}
To allow firmware fuzzers to appropriately solve string comparisons found in many firmware, we developed \splits{}. When integrated with \fuzzware{}~\cite{fuzzware} and AFL++~\cite{aflplusplus}, our instrumentation and mutation techniques were effective in finding test cases suitable for string replacement, and efficiently performing the replacement to reach deeper code. Eleven of the 12 tested firmware containing string comparisons demonstrated statistically significant improvements in block coverage compared to the baseline \fuzzware{} with 6 \textit{new} bugs (with one recent CVE assignment) in firmware found through the inclusion of \splits{}.

%
%
%
\bibliographystyle{splncs04}
\bibliography{references}

\appendix
\section{Appendix}

\subsection{Detailed Performance Analysis}\label{apd:deep-dive}

\begin{table}[t!]
\centering
\caption{Time to solve a sample of string comparisons over five 24 hour fuzzing campaigns. The fastest minimum, median and maximum values for each firmware are shown in \textbf{bold} and gray regions show failures.}
\label{fig:solve_time_tab}
\resizebox{0.83\linewidth}{!}{
\begin{tabular}{|l|l|l|l|l||l|l|l||l|l|l|} 
\cline{3-11}
\multicolumn{1}{c}{} & \multicolumn{1}{c|}{}                & \multicolumn{3}{c||}{\fuzzware{}} & \multicolumn{3}{c||}{\icicle{}  (CompCov)} & \multicolumn{3}{c|}{\splits{}}     \\ 
\hline
\textbf{Firmware}          & \textbf{String}                               & \textbf{Min}    & \textbf{Med} & \textbf{Max}         & \textbf{Min}          & \textbf{Med}      & \textbf{Max}            & \textbf{Min}          & \textbf{Med}      & \textbf{Max} \\ 
\hline
\hline
LiteOS IoT        & OK                                   & 2h13m  & \cellcolor[rgb]{0.8,0.8,0.8}{\textgreater{}24h}    & \cellcolor[rgb]{0.8,0.8,0.8}{\textgreater{}24h}  & 18h23m  & \cellcolor[rgb]{0.8,0.8,0.8}{\textgreater{}24h}    & \cellcolor[rgb]{0.8,0.8,0.8}{\textgreater{}24h}  & \textbf{19m} & \textbf{1h2m} & \textbf{1h32m}  \\ 
\hline
RF Door Lock      & OK\textbackslash{}r\textbackslash{}n & 1h1m & \cellcolor[rgb]{0.8,0.8,0.8}{\textgreater{}24h}    & \cellcolor[rgb]{0.8,0.8,0.8}{\textgreater{}24h}   & 2h53m & 20h28m & \cellcolor[rgb]{0.8,0.8,0.8}{\textgreater{}24h} & \textbf{2m}  & \textbf{4m}  & \textbf{10m}     \\
\hline
\hline
3D Printer          & M108      & \cellcolor[rgb]{0.8,0.8,0.8}{\textgreater{}24h}  & \cellcolor[rgb]{0.8,0.8,0.8}{\textgreater{}24h}  & \cellcolor[rgb]{0.8,0.8,0.8}{\textgreater{}24h} & \cellcolor[rgb]{0.8,0.8,0.8}{\textgreater{}24h}  & \cellcolor[rgb]{0.8,0.8,0.8}{\textgreater{}24h}  & \cellcolor[rgb]{0.8,0.8,0.8}{\textgreater{}24h} & \textbf{11m} & \textbf{23m} & \textbf{44m} \\ 
\hline
3D Printer          & M112      & \cellcolor[rgb]{0.8,0.8,0.8}{\textgreater{}24h}  & \cellcolor[rgb]{0.8,0.8,0.8}{\textgreater{}24h}  & \cellcolor[rgb]{0.8,0.8,0.8}{\textgreater{}24h} & 11h7m  & \cellcolor[rgb]{0.8,0.8,0.8}{\textgreater{}24h}  & \cellcolor[rgb]{0.8,0.8,0.8}{\textgreater{}24h} & \textbf{11m} & \textbf{23m} & \textbf{44m} \\ 
\hline
3D Printer          & M410      & \cellcolor[rgb]{0.8,0.8,0.8}{\textgreater{}24h}  & \cellcolor[rgb]{0.8,0.8,0.8}{\textgreater{}24h}  & \cellcolor[rgb]{0.8,0.8,0.8}{\textgreater{}24h} & \cellcolor[rgb]{0.8,0.8,0.8}{\textgreater{}24h}  & \cellcolor[rgb]{0.8,0.8,0.8}{\textgreater{}24h}  & \cellcolor[rgb]{0.8,0.8,0.8}{\textgreater{}24h} & \textbf{11m} & \textbf{23m} & \textbf{44m} \\ 
\hline
3D Printer          & M110      & \cellcolor[rgb]{0.8,0.8,0.8}{\textgreater{}24h}  & \cellcolor[rgb]{0.8,0.8,0.8}{\textgreater{}24h} & \cellcolor[rgb]{0.8,0.8,0.8}{\textgreater{}24h} & \cellcolor[rgb]{0.8,0.8,0.8}{\textgreater{}24h}  & \cellcolor[rgb]{0.8,0.8,0.8}{\textgreater{}24h}  & \cellcolor[rgb]{0.8,0.8,0.8}{\textgreater{}24h} & \textbf{46m} & \textbf{1h33m} & \textbf{1h59m} \\ 
\hline
Steering Control     & steer     & \cellcolor[rgb]{0.8,0.8,0.8}{\textgreater{}24h}  & \cellcolor[rgb]{0.8,0.8,0.8}{\textgreater{}24h} & \cellcolor[rgb]{0.8,0.8,0.8}{\textgreater{}24h} & \cellcolor[rgb]{0.8,0.8,0.8}{\textgreater{}24h}  & \cellcolor[rgb]{0.8,0.8,0.8}{\textgreater{}24h}  & \cellcolor[rgb]{0.8,0.8,0.8}{\textgreater{}24h} & \textbf{6m}  & \textbf{8m}    & \textbf{13m}  \\ 
\hline
Steering Control     & motor     & \cellcolor[rgb]{0.8,0.8,0.8}{\textgreater{}24h}  & \cellcolor[rgb]{0.8,0.8,0.8}{\textgreater{}24h} & \cellcolor[rgb]{0.8,0.8,0.8}{\textgreater{}24h} & \cellcolor[rgb]{0.8,0.8,0.8}{\textgreater{}24h}  & \cellcolor[rgb]{0.8,0.8,0.8}{\textgreater{}24h}  & \cellcolor[rgb]{0.8,0.8,0.8}{\textgreater{}24h} & \textbf{6m}  & \textbf{8m}    & \textbf{13m} \\
\hline
\hline
Console          & ps               & 3m          & 41m          & 2h22m    & \textbf{2m}          & 24m          & 1h25m      & \textbf{2m}      & \textbf{3m}      & \textbf{7m}  \\ 
\hline
Console          & reboot    & \cellcolor[rgb]{0.8,0.8,0.8}{\textgreater{}24h}  & \cellcolor[rgb]{0.8,0.8,0.8}{\textgreater{}24h}   & \cellcolor[rgb]{0.8,0.8,0.8}{\textgreater{}24h} & 13h47m       & \cellcolor[rgb]{0.8,0.8,0.8}{\textgreater{}24h}  & \cellcolor[rgb]{0.8,0.8,0.8}{\textgreater{}24h} & \textbf{2m}  & \textbf{6m}   & \textbf{7m} \\ 
\hline
Console          & help      & \cellcolor[rgb]{0.8,0.8,0.8}{\textgreater{}24h}  & \cellcolor[rgb]{0.8,0.8,0.8}{\textgreater{}24h}   & \cellcolor[rgb]{0.8,0.8,0.8}{\textgreater{}24h} & 3h40m  & 21h52m  & \cellcolor[rgb]{0.8,0.8,0.8}{\textgreater{}24h} & \textbf{2m}  & \textbf{6m}   & \textbf{6m}  \\ 
\hline
Console          & saul      & \cellcolor[rgb]{0.8,0.8,0.8}{\textgreater{}24h} & \cellcolor[rgb]{0.8,0.8,0.8}{\textgreater{}24h} & \cellcolor[rgb]{0.8,0.8,0.8}{\textgreater{}24h} & 9h3m  & \cellcolor[rgb]{0.8,0.8,0.8}{\textgreater{}24h}   & \cellcolor[rgb]{0.8,0.8,0.8}{\textgreater{}24h} & \textbf{2m} & \textbf{6m} & \textbf{7m}  \\ 
\hline
Console     & write          & \cellcolor[rgb]{0.8,0.8,0.8}{\textgreater{}24h} & \cellcolor[rgb]{0.8,0.8,0.8}{\textgreater{}24h} & \cellcolor[rgb]{0.8,0.8,0.8}{\textgreater{}24h} & \cellcolor[rgb]{0.8,0.8,0.8}{\textgreater{}24h}  & \cellcolor[rgb]{0.8,0.8,0.8}{\textgreater{}24h}   & \cellcolor[rgb]{0.8,0.8,0.8}{\textgreater{}24h} & \textbf{54m} & \textbf{1h25m}  & \textbf{4h25m}  \\ 
\hline
Console     & read           & \cellcolor[rgb]{0.8,0.8,0.8}{\textgreater{}24h} & \cellcolor[rgb]{0.8,0.8,0.8}{\textgreater{}24h} & \cellcolor[rgb]{0.8,0.8,0.8}{\textgreater{}24h} & \cellcolor[rgb]{0.8,0.8,0.8}{\textgreater{}24h}  & \cellcolor[rgb]{0.8,0.8,0.8}{\textgreater{}24h}   & \cellcolor[rgb]{0.8,0.8,0.8}{\textgreater{}24h} & \textbf{51m}  & \textbf{1h4m} & \textbf{3h22m} \\
\hline
Console     & all            & \cellcolor[rgb]{0.8,0.8,0.8}{\textgreater{}24h} & \cellcolor[rgb]{0.8,0.8,0.8}{\textgreater{}24h} & \cellcolor[rgb]{0.8,0.8,0.8}{\textgreater{}24h} & \cellcolor[rgb]{0.8,0.8,0.8}{\textgreater{}24h}  & \cellcolor[rgb]{0.8,0.8,0.8}{\textgreater{}24h}   & \cellcolor[rgb]{0.8,0.8,0.8}{\textgreater{}24h} & \textbf{1h5m}  & \textbf{1h14m} & \textbf{4h19m} \\
\hline
Console     & rtc            & 14h30m            & \cellcolor[rgb]{0.8,0.8,0.8}{\textgreater{}24h} & \cellcolor[rgb]{0.8,0.8,0.8}{\textgreater{}24h} & 4h58m  & \cellcolor[rgb]{0.8,0.8,0.8}{\textgreater{}24h}   & \cellcolor[rgb]{0.8,0.8,0.8}{\textgreater{}24h} & \textbf{2m}    & \textbf{3m}   & \textbf{7m} \\
\hline
Console     & poweron        & \cellcolor[rgb]{0.8,0.8,0.8}{\textgreater{}24h} & \cellcolor[rgb]{0.8,0.8,0.8}{\textgreater{}24h} & \cellcolor[rgb]{0.8,0.8,0.8}{\textgreater{}24h} & 9h25m  & \cellcolor[rgb]{0.8,0.8,0.8}{\textgreater{}24h}   & \cellcolor[rgb]{0.8,0.8,0.8}{\textgreater{}24h} & \textbf{7m}   & \textbf{1h26m}   & \textbf{1h40m} \\
\hline
Console     & poweroff       & \cellcolor[rgb]{0.8,0.8,0.8}{\textgreater{}24h} & \cellcolor[rgb]{0.8,0.8,0.8}{\textgreater{}24h} & \cellcolor[rgb]{0.8,0.8,0.8}{\textgreater{}24h} & 9h38m  & \cellcolor[rgb]{0.8,0.8,0.8}{\textgreater{}24h}   & \cellcolor[rgb]{0.8,0.8,0.8}{\textgreater{}24h} & \textbf{7m}   & \textbf{1h26m}   & \textbf{1h40m} \\
\hline
Console     & clearalarm     & \cellcolor[rgb]{0.8,0.8,0.8}{\textgreater{}24h} & \cellcolor[rgb]{0.8,0.8,0.8}{\textgreater{}24h} & \cellcolor[rgb]{0.8,0.8,0.8}{\textgreater{}24h} & \cellcolor[rgb]{0.8,0.8,0.8}{\textgreater{}24h}  & \cellcolor[rgb]{0.8,0.8,0.8}{\textgreater{}24h}   & \cellcolor[rgb]{0.8,0.8,0.8}{\textgreater{}24h} & \textbf{7m}   & \textbf{1h26m}   & \textbf{1h40m} \\
\hline
Console     & getalarm       & \cellcolor[rgb]{0.8,0.8,0.8}{\textgreater{}24h} & \cellcolor[rgb]{0.8,0.8,0.8}{\textgreater{}24h} & \cellcolor[rgb]{0.8,0.8,0.8}{\textgreater{}24h} & 18h40m  & \cellcolor[rgb]{0.8,0.8,0.8}{\textgreater{}24h}   & \cellcolor[rgb]{0.8,0.8,0.8}{\textgreater{}24h} & \textbf{7m}   & \textbf{1h26m}   & \textbf{1h40m} \\
\hline
Console     & setalarm       & \cellcolor[rgb]{0.8,0.8,0.8}{\textgreater{}24h} & \cellcolor[rgb]{0.8,0.8,0.8}{\textgreater{}24h} & \cellcolor[rgb]{0.8,0.8,0.8}{\textgreater{}24h} & 20h37m  & \cellcolor[rgb]{0.8,0.8,0.8}{\textgreater{}24h}   & \cellcolor[rgb]{0.8,0.8,0.8}{\textgreater{}24h} & \textbf{31m}   & \textbf{1h48m} & \textbf{4h28m} \\
\hline
Console     & gettime        & \cellcolor[rgb]{0.8,0.8,0.8}{\textgreater{}24h} & \cellcolor[rgb]{0.8,0.8,0.8}{\textgreater{}24h} & \cellcolor[rgb]{0.8,0.8,0.8}{\textgreater{}24h} & \cellcolor[rgb]{0.8,0.8,0.8}{\textgreater{}24h}  & \cellcolor[rgb]{0.8,0.8,0.8}{\textgreater{}24h}   & \cellcolor[rgb]{0.8,0.8,0.8}{\textgreater{}24h} & \textbf{7m}   & \textbf{1h25m} & \textbf{1h39m}  \\
\hline
Console     & settime        & \cellcolor[rgb]{0.8,0.8,0.8}{\textgreater{}24h} & \cellcolor[rgb]{0.8,0.8,0.8}{\textgreater{}24h} & \cellcolor[rgb]{0.8,0.8,0.8}{\textgreater{}24h} & \cellcolor[rgb]{0.8,0.8,0.8}{\textgreater{}24h}  & \cellcolor[rgb]{0.8,0.8,0.8}{\textgreater{}24h}   & \cellcolor[rgb]{0.8,0.8,0.8}{\textgreater{}24h} & \textbf{31m}   & \textbf{2h15m} & \textbf{3h17m} \\
\hline
\end{tabular}
}
\end{table}

For a deeper analysis of the effectiveness of our approach, we selected three sets of firmware based on their use of strings to determine the efficiency with which comparisons are solved by \splits{} compared to the state-of-the-art fuzzers.

As shown in Table~\ref{fig:solve_time_tab}, \splits{} was able to quickly and consistently solve the error checks in the LiteOS IoT and RF Door Lock binaries. Neither \fuzzware{} or \icicle{} with CompCov could consistently solve these comparisons. For \fuzzware{}, only a single run for each binary was able to pass this comparison and reach the main loop. In our 3D Printer and Steering Control tests, using strings to receive data, \splits{} solved all of these comparisons. In cases where multiple, similar length, strings are compared in quick succession, \splits{} would solve each of these strings within seconds of each other. \fuzzware{} did not solve any of these comparisons, while \icicle{} with CompCov only solved a single string from this set. For the console binary, with \splits{}, all string comparisons were solved in every run, while \fuzzware{} alone solved the shortest string, \texttt{ps} consistently, and the second shortest string \texttt{rtc} in a single run. \icicle{} with CompCov solved more strings than \fuzzware{} due to CompCov instrumentation, but still lacked consistency and the process took considerably longer than \splits{}.

\subsection{Code Coverage (Magic-String-Binaries and Other-Binaries)}\label{apd:coverage}
Table~\ref{fig:pval_tab} and Table \ref{tab:nostringcov} show the coverage achieved in the \textit{Magic-String-Binaries} and \textit{Other-Binaries} respectively.

\begin{table}[h]
\centering
\vspace{-4mm}
\caption{\textit{Magic-String-Binaries.} Fuzzing minimum, median and maximum block coverage achieved with each fuzzing framework over five 24 hour fuzzing campaigns. P-Values indicating statistical significance are calculated using Mann Whitney U tests, conducted at a 0.01 significance level. The highest minimum, median and maximum values for each firmware are shown in \textbf{bold}.}
\label{fig:pval_tab}
\resizebox{\linewidth}{!}{
\begin{tabular}{|l|l|l|l|l||l|l|l||l|l|l||l|l|l|l|l|l|} 
\cline{3-17}
\multicolumn{1}{c}{} & \multicolumn{1}{c|}{}                                          & \multicolumn{3}{c||}{\fuzzware{}} & \multicolumn{3}{c||}{\ember{}} & \multicolumn{3}{c||}{\icicle{} (CompCov)}                & \multicolumn{6}{c|}{\splits{}}                \\ 
\hline
\textbf{Firmware}             & \begin{tabular}[c]{@{}l@{}}\textbf{Blocks in}\\\textbf{Firmware}\end{tabular} & \textbf{Min}           & \textbf{Med}        & \textbf{Max} & \textbf{Min}           & \textbf{Med}        & \textbf{Max} & \textbf{Min}           & \textbf{Med}        & \textbf{Max}          & \textbf{Min}           & \textbf{Med}        & \textbf{Max}          & \begin{tabular}[c]{@{}l@{}}\textbf{p-val to}\\\textbf{\fuzzware{}}\end{tabular} & \begin{tabular}[c]{@{}l@{}}\textbf{p-val to}\\\textbf{\ember{}}\end{tabular} & \begin{tabular}[c]{@{}l@{}}\textbf{p-val to}\\\textbf{\icicle{}}\end{tabular} \\ 
\hline
\hline
3D Printer          & 8045    & 1229   & 1289   &  1383 & 1575 & \textbf{3517} & \textbf{4059} & 1311   & 1445   &  2988  & \textbf{2695}   & 3134  & 3614  &  \textless0.01  & 0.465  & 0.016\\ 
\hline
Console              & 2251    & 803    & 805    & 844 & 804 & 843 & 856  & 808   & 830   &  1063  & \textbf{1157}   & \textbf{1160}  & \textbf{1161}  &  \textless0.01  & \textless0.01 & \textless0.01\\ 
\hline
GPS Tracker          & 4194    & 747    & 748   & 754 & 756 & 756 & 759 & 748   & 750   &  753  & \textbf{827}   & \textbf{830}  & \textbf{833}  &  \textless0.01  & \textless0.01 & \textless0.01\\
\hline
LiteOS IoT           & 2423    & 736    & 737    & 1346 & 1347 & 1348 & 1350 & 736   & 736   &  1313  & \textbf{1358}   & \textbf{1366}  & \textbf{1368}  &  \textless0.01 & \textless0.01 & \textless0.01 \\ 
\hline
RF Door Lock         & 3320    & 781    & 782    & 2548 & 782 & 782 & 2662 & 780   & 1177   &  2380  & \textbf{2539}   & \textbf{2553}  & \textbf{2687}  &  0.015  & 0.067 & \textless0.01\\ 
\hline
Steering Control     & 1835    & 609    & 613    & 620 & 638 & 644 & 648 & 610   & 613   &  615   & \textbf{643}    & \textbf{648}   & \textbf{652}   &  \textless0.01  & 0.169 & \textless0.01 \\ 
\hline
uTasker MODBUS       & 3780    & 1244   & 1246   & 1280 & 1219 & 1252 & 1311 & 1246   & 1303   &  1303   & \textbf{2042}   & \textbf{2052}  & \textbf{2100}  &  \textless0.01  & \textless0.01 & \textless0.01\\ 
\hline
uTasker USB         & 3491    & 1734   & 1745   & 1775 & 1341 & 1351 & 1360 & 1520   & 1540   &  1862   & \textbf{1792}   & \textbf{1815}  & \textbf{1934}  &  \textless0.01   & \textless0.01 & 0.047\\ 
\hline
Zephyr SocketCan     & 5943    & 2689   & 2976   & 3029 & 2272 & 2468 & 2565 & 2733   & 2806   &  2809  & \textbf{3093}   & \textbf{3126}  & \textbf{3135}  &  \textless0.01 & \textless0.01 & \textless0.01 \\ 
\hline
ChibiOS RTC     & 3013    & 559   & 578   & 593 & 554 & 558 & 565 & 554   & 567   &  575  & \textbf{1002}   & \textbf{1005}  & \textbf{1007}  &  \textless0.01  & \textless0.01 & \textless0.01 \\ 
\hline
Contiki-NG Shell     & 4776    & 1593   & 1594   & 1596 & 1564 & 1595 & 1596 & 1584   & 1587   &  1590  & \textbf{1874}   & \textbf{1973}  & \textbf{1993}  &  \textless0.01   & \textless0.01 & \textless0.01\\ 
\hline
RiotOS TWR     & 4261    & 1219   & 1222   & 1224 & 593 & 593 & 1224  & 1208   & 1218   &  1291 & \textbf{1415}   & \textbf{1436}  & \textbf{1618}  &  \textless0.01   & \textless0.01 & \textless0.01 \\ 
\hline
\end{tabular}
}
\vspace{-8mm}
\end{table}

\begin{table}[h]
\centering
\caption{\textit{Other-Binaries.~}Minimum (min), median (med), maximum (max) block coverage achieved with each fuzzing framework over five 24 hour trials. P-values indicating statistical significance are calculated using Mann Whitney U tests, at a 0.01 significance level. No binaries showed a statistically significant difference. The highest \textit{min}, \textit{med} and \textit{max} values for each firmware are shown in \textbf{bold}.}
\label{tab:nostringcov}
\resizebox{0.68\linewidth}{!}{
\begin{tabular}{|l|l|l|l|l||l|l|l|l|} 
\cline{3-9}
\multicolumn{1}{c}{} & \multicolumn{1}{c|}{}                                          & \multicolumn{3}{c||}{\textsc{Fuzzware}} & \multicolumn{4}{c|}{\splits{}}  \\ 
\hline
\textbf{Firmware}             & \begin{tabular}[c]{@{}l@{}}\textbf{Blocks in}\\\textbf{Firmware}\end{tabular} & \textbf{Min}           & \textbf{Med}        & \textbf{Max}           & \textbf{Min}           & \textbf{Med}        & \textbf{Max}          & \textbf{p-value}  \\  
\hline
\hline
6LoWPAN Receiver     & 6977   & 2732 & \textbf{3149}   & \textbf{3206}          & \textbf{2830} & 3056   & 3119 &  0.175                   \\ 
\hline
6LoWPAN Sender       & 6980   & 2772 & 2972   & 3161          & \textbf{2786} & \textbf{3113}   & \textbf{3273} &  0.347                   \\ 
\hline
CNC                  & 3614   & \textbf{2561} & \textbf{2718}   & \textbf{2733}          & 2209 & 2510   & 2611 &   0.016                  \\ 
\hline
Drone                & 2728   & \textbf{1826} & \textbf{1828}   & \textbf{1843}          & 713 & 1734   & 1837 &  0.076                   \\ 
\hline
Gateway              & 4921   & \textbf{2908} & \textbf{2939}   & \textbf{3127}          & 2408 & 2686   & 2939 &  0.036                   \\ 
\hline
Heat Press           & 1837   & \textbf{550}  & \textbf{554}    & \textbf{564}           & 549  & 550    & 556  &  0.164                   \\ 
\hline
PLC                  & 2303   & \textbf{637}  & \textbf{644}    & \textbf{907}           & 629  & 642    & 650  &  0.344                   \\ 
\hline
Reflow Oven          & 2947   & \textbf{1192} & \textbf{1192}   & \textbf{1192}          & \textbf{1192} & \textbf{1192}   & \textbf{1192} &  N/A                   \\ 
\hline
Robot                & 3034   & \textbf{1305} & \textbf{1313}   & 1315          & 1298 & 1306   & \textbf{1319} &  0.249                   \\ 
\hline
Soldering Iron       & 3656   & \textbf{2302} & 2353   & 2457          & 2229 & \textbf{2457}   & \textbf{2465} &  0.528                   \\ 
\hline
Thermostat           & 4673   & 3245 & 3410   & 3497          & \textbf{3308} & \textbf{3430}   & \textbf{3504} &  0.251                   \\ 
\hline
XML Parser           & 9376   & 3239 & 3634   & 3826          & \textbf{3418} & \textbf{3850}   & \textbf{4004} &  0.175                   \\
\hline
\end{tabular}
}
\end{table}

\end{document}